\documentclass[a4paper,review,11pt,authoryear]{elsarticle}
\usepackage{kpfonts}
\usepackage[table]{xcolor}
\usepackage{kpfonts}
\usepackage{amsmath,bm,hyperref,url,fullpage,mathtools,booktabs,parskip}
\usepackage{paralist,enumitem,todonotes,microtype}
\usepackage[a4paper,text={16cm,25cm}]{geometry}
\usepackage[labelsep=period]{caption}

\usepackage[section]{placeins}
\usepackage{algorithm}
\usepackage{algorithmicx}
\usepackage{algpseudocode}
\usepackage{verbatim}
\usepackage{multirow}
\usepackage{multicol}
\usepackage{arydshln}
\usepackage{color}
\usepackage{adjustbox}
\usepackage{amssymb}
\usepackage{listings}

\usepackage[toc,page]{appendix}

\usepackage{makecell}
\usepackage{bbm}
\usepackage{dsfont}

\providecommand{\tightlist}{
  \setlength{\itemsep}{0pt}\setlength{\parskip}{0pt}}

\graphicspath{{IJFpaper-figure/}}

\usepackage{hyperref}
\hypersetup{
colorlinks=true,
linkcolor=blue,
citecolor=blue,
urlcolor=blue}

\let\code=\texttt
\let\proglang=\textsf
\newcommand{\pkg}[1]{{\fontseries{b}\selectfont #1}}
\DeclareMathOperator{\Fscore}{F_{1}-score}

\newcommand\Algphase[1]



\journal{International Journal of Forecasting}
\begin{document}
\date{}
\begin{frontmatter}

\title{fETSmcs: Feature-based ETS model component selection}

\author[mainaddress]{Lingzhi Qi\fnref{eqc}}\ead{lingzhi_qi@163.com}
    \author[thirdaddress]{Xixi Li\fnref{eqc}\corref{cor}}\ead{xixi.li@manchester.ac.uk} \ead[url]{https://orcid.org/0000-0001-5846-3460}
 \author[mainaddress]{Qiang Wang} \ead{wang6965@sina.com}
 \author[mainaddress]{Suling Jia} \ead{jiasuling@126.com}
 \address[mainaddress]{School of Economics and Management,
 	Beihang University, China.}
 \address[thirdaddress]{Department of Mathematics, University of Manchester, UK.}
 \fntext[eqc]{The authors contributed equally.}
   \cortext[cor]{Corresponding author}

\begin{abstract}
The well-developed ETS (ExponenTial Smoothing or Error, Trend, Seasonality) method incorporating a family of exponential smoothing models in state space representation has been widely used for automatic forecasting. The existing ETS method uses information criteria for model selection by choosing an optimal model with the smallest information criterion among all models fitted to a given time series. The ETS method under such a model selection scheme suffers from computational complexity when applied to large-scale time series data. To tackle this issue, we propose an efficient approach for ETS model selection by training classifiers on simulated data to predict appropriate model component forms for a given time series. We provide a simulation study to show the model selection ability of the proposed approach on simulated data. We evaluate our approach on the widely used forecasting competition data set M4, in terms of both point forecasts and prediction intervals. To demonstrate the practical value of our method, we showcase the performance improvements from our approach on a monthly hospital data set.

\end{abstract}


\end{frontmatter}

\section{Introduction}
\label{introduction}

In today's data-rich environment, modern business decisions often involve forecasting large collections of time series data, and people often want to select optimal models for their data in an effective and automatic manner \citep{petropoulos2022forecasting}. However, selecting one optimal model requires professional knowledge and relevant experience, making automatic and accurate forecasting not a straightforward task \citep{li2022improving}.
 
 \cite{hyndman2002state} incorporated a family of exponential smoothing models into a well-developed modeling framework namely ETS for automatic forecasting. The ETS method is able to identify three components of a given time series: error, trend, and seasonality, as well as their component forms (additive or multiplicative). The error component can be identified as additive (`A') or multiplicative (`M'). The trend and seasonality components can be none (`N'), `A' or `M'. Also, the trend can be damped or not.

In the context of ETS modeling, model selection refers to identifying appropriate forms of three components (error, trend and seasonality) for a given time series.  The ETS method uses information criteria such as AIC (Akaike Information Criterion), AICc (The corrected Akaike Information Criterion), and BIC (Bayesian Information Criterion) for the selection purpose, and it proceeds in two steps: (i) calculate  information criteria of all candidate models fitted to data; and (ii) select a model with the smallest information criterion among all models.  However, in modern business environment, large retailers often need to forecast large-scale data from thousands of products in a forecast cycle \citep{li2022improving}. For example, Walmart has roughly 5,000 stores in the U.S., and it is required to generate approximately 1 billion unique forecasts for business decisions \citep{seaman2018considerations}. The existing ETS method under such a model selection scheme is computationally expensive when used to forecast large-scale time series data. There is an urgent need to propose an efficient approach for ETS model selection to forecast large collections of time series in a batch and effective manner. 

The issue of the computation cost of a large family of models has attracted some attention in recent studies. \cite{nikolopoulos2018forecasting} proposed suboptimal forecasting solutions for faster and robust systems where they compared the performance of a suboptimal search for smoothing parameters of exponential smoothing models with that of a more computationally expensive and optimal search.
\cite{petropoulos2021fast} also questioned the necessity of such a large family of models, and suggested a change from a focus on the search of suboptimal parameters to that of suboptimal models.

A data-driven approach that uses time series features for forecast model selection and combination over large-scale time series data has flourished in recent studies \citep[e.g.,][]{montero2020fforma,li2020forecasting,talagala2021fformpp}. However, the interest of existing feature-based model selection focused more on the selection among different types of models. Until now, the use of features for model selection among a family of models has not yet attracted enough attention.  In order to improve the efficiency of the ETS method, we explore the possibility of the use of features for ETS model selection by training classifiers on simulated data to predict model component forms for a given time series.

We show performance improvements in model selection from our method on simulated data.
We conduct extensive empirical evaluations of the proposed method by applying it to 95 thousand real time series from the Makridakis forecasting competitions \citep{Makridakis2020-mm}. 
Significance tests results show that the proposed approach results in robust forecasts at long-term and all horizons for yearly, quarterly and monthly data. 
To demonstrate the practical value of the proposed approach, we show the performance improvements from our method on a hospital data set compared with the information criteria method.

The key innovations of our approach are as follows.
\begin{itemize}
 \tightlist
\item The existing use of features for model selection focused more on the selection among different types of models while our work makes the first attempt to select an appropriate model among a family of models.
\item The proposed ETS model selection approach can effectively reduce computational cost when applied to large-scale time series data by predicting model component forms using pre-trained classifiers.
 
\end{itemize}

The rest of the article is organized as follows: Section~\ref{background} offers a short literature review on relevant studies.
Section~\ref{method} describes the methodology for the proposed approach. Section~\ref{experiments} presents the empirical results including a simulation study, real data applications and significance tests.
Section~\ref{case-study} offers a case study.
Section~\ref{discussion} offers our discussions and insights. Finally, Section~\ref{conclusion} provides our conclusions.

\section{Background research}
\label{background}
\subsection{The exponential smoothing family of models}
The exponential smoothing methods have been around since the 1950s when \cite{brown1956exponential} proposed a simple exponential smoothing method for  demand prediction.
The work of \cite{brown1959statistical} and \cite{gardner1985exponential} led to the development of exponential smoothing models towards automated forecasting \citep{hyndman2002state}. 
 \cite{ord1997estimation} and \cite{hyndman2002state} showed that all exponential smoothing models can be well written in the form of state space representation.  \cite{hyndman2002state} incorporated a family of exponential smoothing models into a well-developed modeling framework namely ETS. The framework incorporates state space models, parameter estimation, point prediction and interval prediction.

\subsection{Information criteria for model selection}

Information criteria approaches are designed for parametric methods under the framework of maximum likelihood estimation. They are used to correct for the bias of maximum likelihood by adding a penalty term to compensate for the over-fitting of more complex models \citep{bishop2006pattern}. 
These commonly used information criteria are employed to evaluate how well a given model fits the data, and how complicated the fitted model is.

Although information criteria could partially address the problem of over-fitting \citep{meira2021treating}, the fact that the information criteria are based on one-step prediction error implies that the selected model does not necessarily produce accurate and reliable forecasts when applied to long forecasting horizons \citep{fildes2015simple}. To tackle this issue, cross-validation can be used to assess the performance of models over multiple time windows, but this is not necessarily guaranteed, and \cite{billah2006exponential} showed that the performance of information criteria approaches is superior to that of the commonly used validation approach. However, the use of cross-validation is suitable for longer series and is computationally expensive \citep{li2022improving}. What's more, the information criteria approaches do not take parameters uncertainty into consideration \citep{bishop2006pattern}, and in practice, they tend to select simple models.

\subsection{Feature-based model selection and combination}
The use of machine learning for model selection is not novel, many attempts have been made at feature-based forecasting for univariate time series. \cite{reid1972comparison} pointed out that the forecasting performance is influenced by time series features. \cite{kang2017visualising} projected time series into an instance space, and provided some insights into the forecasting performance of different models.
Recently, time series features have been used for forecast model selection and combination, aiming at selecting an appropriate model or predicting combination weights of all candidate models for a given time series. The performance of feature-based forecasting is highly dependent on the selection of an appropriate set of time series features. \cite{collopy1992rule} developed a rule-based system to select optimal models for the series based on 99 features. \cite{talagala2018meta} trained a decision tree for the model selection purpose using 42 features. \cite{montero2020fforma} trained XGBoost to predict an optimal combination weight of each candidate model. Their method ranked 2nd in M4 competition \citep{Makridakis2020-mm}. \cite{talagala2021fformpp} predicted forecasting error using time series features with a Bayesian multivariate surface regression model. 
Instead of using manual features, \cite{li2020forecasting} proposed an automated approach to extracting time series imaging features, and the extracted features are also used for forecast model combination. 

\section{Feature-based ETS model component selection}
\label{method}

\subsection{Taxonomy of exponential smoothing models in the ETS method}
Table~\ref{tab:etsm} presents the taxonomy of exponential smoothing models in the ETS method.  Theoretically, the number of exponential smoothing models from the combinations of three component forms can reach 30, as shown in Table~\ref{tab:etsm}. However, in practice, the 11 models illustrated in Table~\ref{tab:etsm} with gray background would lead to infinite forecast variance problems when applied to long forecasting horizons. For more details about this issue, see Chapter 15 of \cite{hyndman2008forecasting}. Considering that models with multiplicative trend would lead to unrealistically explosive forecasts \citep{petropoulos2021fast}, the \proglang{R} package \pkg{forecast}~\citep{Rforecast} has, by default, excluded these four models (`MMN', `MMdN', `MMM', and `MMdM') by setting the argument \code{allow.multiplicative.trend} as \code{False} in the function \code{ets()}. Hence, the number of applicable models shaded with purple in Table~\ref{tab:etsm} in the ETS method is 15. For non-seasonal data (e.g., yearly data), the number of applicable models is six.

\begin{table}[htbp] 
  \centering
  \caption{Taxonomy of exponential smoothing models in the ETS framework \citep{hyndman2008forecasting}. The 15 applicable models are shaded with purple. The 11 models that would lead to infinite forecast variance problems are shaded with gray. The four models with multiplicative trends that would lead to unrealistically explosive forecasts are shaded with pink. }
  \label{tab:etsm}
  	\scalebox{0.85}{
\begin{tabular}{rrrrrrr}
\toprule
&\multicolumn{3}{c}{Additive \textbf{E}rror}& \multicolumn{3}{c}{Multiplicative \textbf{E}rror}        \\ 
\cmidrule(lr){2-4} \cmidrule(lr){5-7}
& \multicolumn{3}{c}{\textbf{S}easonality}& \multicolumn{3}{c}{\textbf{S}easonality}        \\ 
\cmidrule(lr){2-4} \cmidrule(lr){5-7}
\multicolumn{1}{c}{\textbf{T}rend} & N & A & M & N & A & M\\ 
\cmidrule(lr){1-1}\cmidrule(lr){2-7}
N (None)           &\cellcolor{blue!20}ANN & \cellcolor{blue!20}ANA &\cellcolor{gray!20}ANM & \cellcolor{blue!20}MNN & \cellcolor{blue!20}MNA & \cellcolor{blue!20}MNM\\ 
A (Additive)       & \cellcolor{blue!20}AAN & \cellcolor{blue!20}AAA &\cellcolor{gray!20}AAM & \cellcolor{blue!20}MAN & \cellcolor{blue!20}MAA & \cellcolor{blue!20}MAM\\ 
Ad (Additive damped)       & \cellcolor{blue!20}AAdN & \cellcolor{blue!20}AAdA &\cellcolor{gray!20}AAdM & \cellcolor{blue!20}MAdN & \cellcolor{blue!20}MAdA & \cellcolor{blue!20}MAdM\\ 
M (Multiplicative) & \cellcolor{gray!20}AMN & \cellcolor{gray!20}AMA & \cellcolor{gray!20}AMM &\cellcolor{purple!20}MMN &\cellcolor{gray!20}MMA &\cellcolor{purple!20}MMM\\ 
Md (Multiplicative damped) &\cellcolor{gray!20}AMdN &\cellcolor{gray!20}AMdA &\cellcolor{gray!20}AMdM &\cellcolor{purple!20}MMdN &\cellcolor{gray!20}MMdA &\cellcolor{purple!20}MMdM\\ 
\bottomrule
\end{tabular}
}
\end{table}
\FloatBarrier

\subsection{The proposed framework}
The aim of this work is to employ machine learning methodology to predict an optimal ETS model for a given time series. A simple approach to this task would be to view it as a multi-label classification problem, that is, we train a classifier to link time series features with the 15 applicable ETS models in Tabel~\ref{tab:etsm}.
However, training a classifier with 15 classes is not an easy task. To tackle this issue, an alternative way is to adopt a divide and conquer strategy \citep{zhihua2021machinelearning} that trains a classifier to predict each component form separately.

\begin{figure}[h!]
  \centering
  \includegraphics[width=16cm,height=8cm]{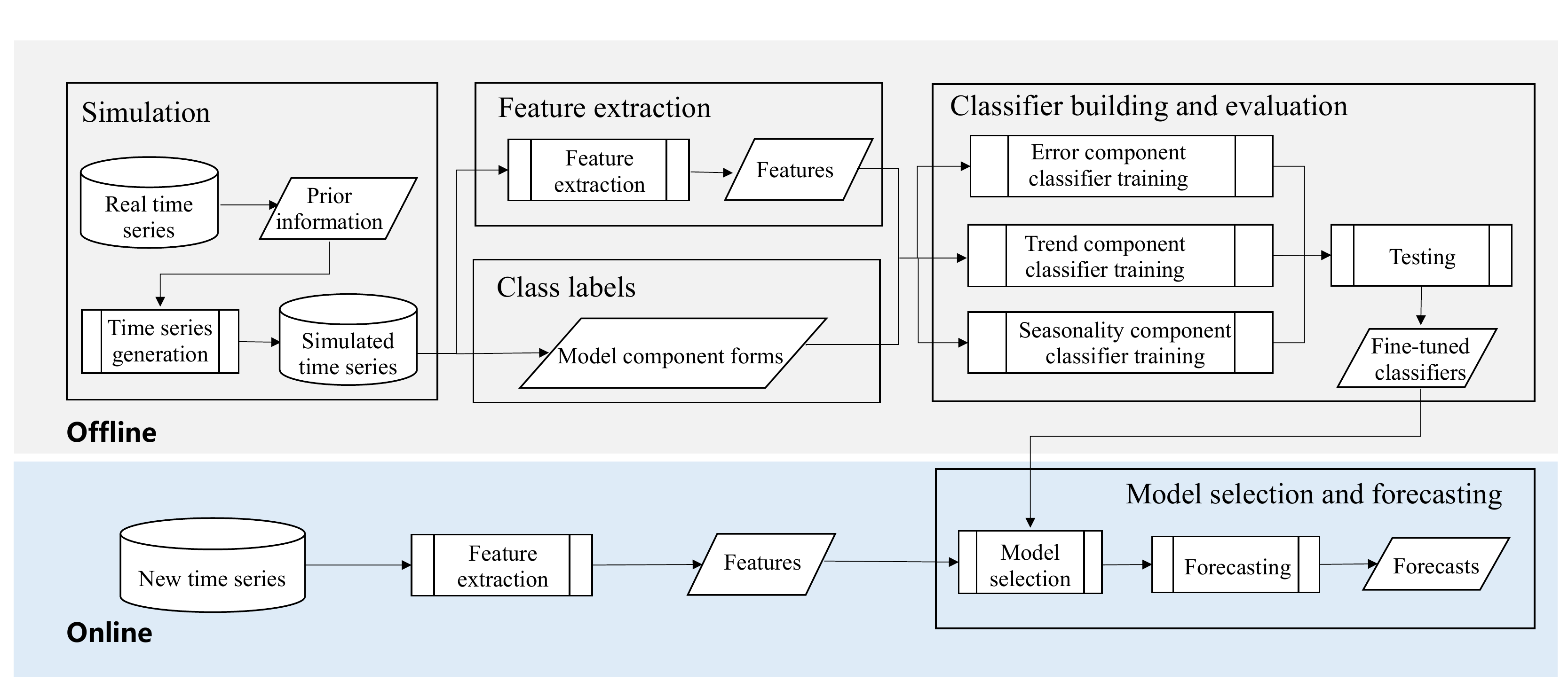}
  \caption{The proposed framework of feature-based ETS model component selection.}
  \label{framework}
\end{figure}

Fig.~\ref{framework} shows the proposed framework of feature-based ETS model selection consisting of the following two parts:
\begin{enumerate}
\def\labelenumi{\arabic{enumi}.}
\tightlist
	\item  \textbf{Offline (Training phase).} The training phase consists of the process of data simulation, feature extraction as well as classifier building and evaluation. In the simulation process, we will simulate some time series from the 15 applicable ETS models. The extracted features over the simulated series with corresponding model component forms will be used for the classifier building and evaluation. Three classifiers will be trained on simulated data separately for the prediction of error, trend and seasonality component forms.

	\item  \textbf{Online (Testing phase).}  The testing part requires feature extraction over any new time series. These extracted features are used as input to the pre-trained classifiers from the training phase to obtain error, trend and seasonality component forms.
	We will check the applicability of a selected model from the combination of predicted component forms, and model adjustment will be performed depending on whether the selected model is applicable.
	Finally, we fit the selected model to data, and then make predictions from the fitted model using the function \code{forecast::ets()} \citep{Rforecast} with the argument \code{model}  being the selected model form. 
\end{enumerate}
\FloatBarrier
\FloatBarrier

\subsection{Feature extraction}
\label{feature-extraction}

 A collection of 42 time series features developed by \cite{montero2020fforma} has been used for forecast model selection and combination over large collections of time series with good results in recent studies \citep[e.g.,][]{montero2020fforma, wang2021uncertainty, talagala2021fformpp}. The combination method of \cite{montero2020fforma} ranked 2nd in one of largest time series competitions namely M4 competition \citep{Makridakis2020-mm}. The collection of features is available in the \proglang{R} package \pkg{M4metalearning} \citep{montero2020fforma}.

Apart from the above features used for forecasting, a collection of 22 high-performing features carefully selected from 7,000 features in the \pkg{hctsa} toolbox \citep{fulcher2017hctsa} has exhibited strong classification performance across different time series problems \citep{lubba2019catch22}. The collection of features is available in the \proglang{R} package \pkg{Rcatch22} \citep{Rcatch22}.
In this work, in order to characterize an observed time series more comprehensively, and thus improve the classification accuracy of model selection, we choose 59 features from these two classes of features for feature extraction. A set of five features from \cite{montero2020fforma} using estimated parameters of fitted ETS models to characterize a time series is discarded. The specific features as well as their corresponding descriptions can be found in Appendix~\ref{features}.

 \subsection{Classifier building for the prediction of ETS model component forms using LightGBM}
 XGBoost and pGBRT are effective implementations of a popular machine learning algorithm GBDT (Gradient Boosting Decision Tree) with many engineering optimizations. Nonetheless, the efficiency and scalability of both are still unsatisfactory when applied to data with high feature dimension and large size \citep{ke2017lightgbm}. To tackle this issue, \cite{ke2017lightgbm} proposed a new GBDT implementation namely LightGBM with two novel techniques GOSS (Gradient-based One-Side Sampling) and EFB (Exclusive Feature Bundling). With GOSS, only small data with large gradients are used for the estimation of information gain. With EFB, the number of features is reduced by bundling mutually exclusive features.

In this study, instead of training a classifier with 15 classes using LightGBM,
we separately train three classifiers $f_{e},f_{t}$ and $f_{s}$ to link time series features with each model component form. 
The three classifiers $f_{e},f_{t}$ and $f_{s}$ for the prediction of error, trend and seasonality component forms are shown in Table \ref{tab:PME}.
 \begin{table}[htbp]
  \centering
  \caption{Three classifiers used for the prediction of three component forms of an ETS model for a given time series.}
  	\label{alg:classification}
  		\scalebox{0.95}{
\begin{tabular}{p{0.08\columnwidth}p{0.6\columnwidth}p{0.1\columnwidth}p{0.15\columnwidth}}
    \toprule
    Classifier &Description&Number of classes&Class labels\\
    \midrule
    $f_{e}$ &A classifier for the prediction of error component form&2 &`A', `M'\\
     $f_{t}$&A classifier for the prediction of trend component form& 3&`A', `Ad', `N'\\
     $f_{s}$&A classifier for the prediction of seasonality component form&3&`A', `M', `N'\\
    \bottomrule
    \end{tabular}
    }
  \label{tab:PME}
\end{table}
\FloatBarrier

\subsection{Model check and adjustment}

\begin{figure}[h!]
  \centering
  \includegraphics[width=1\linewidth]{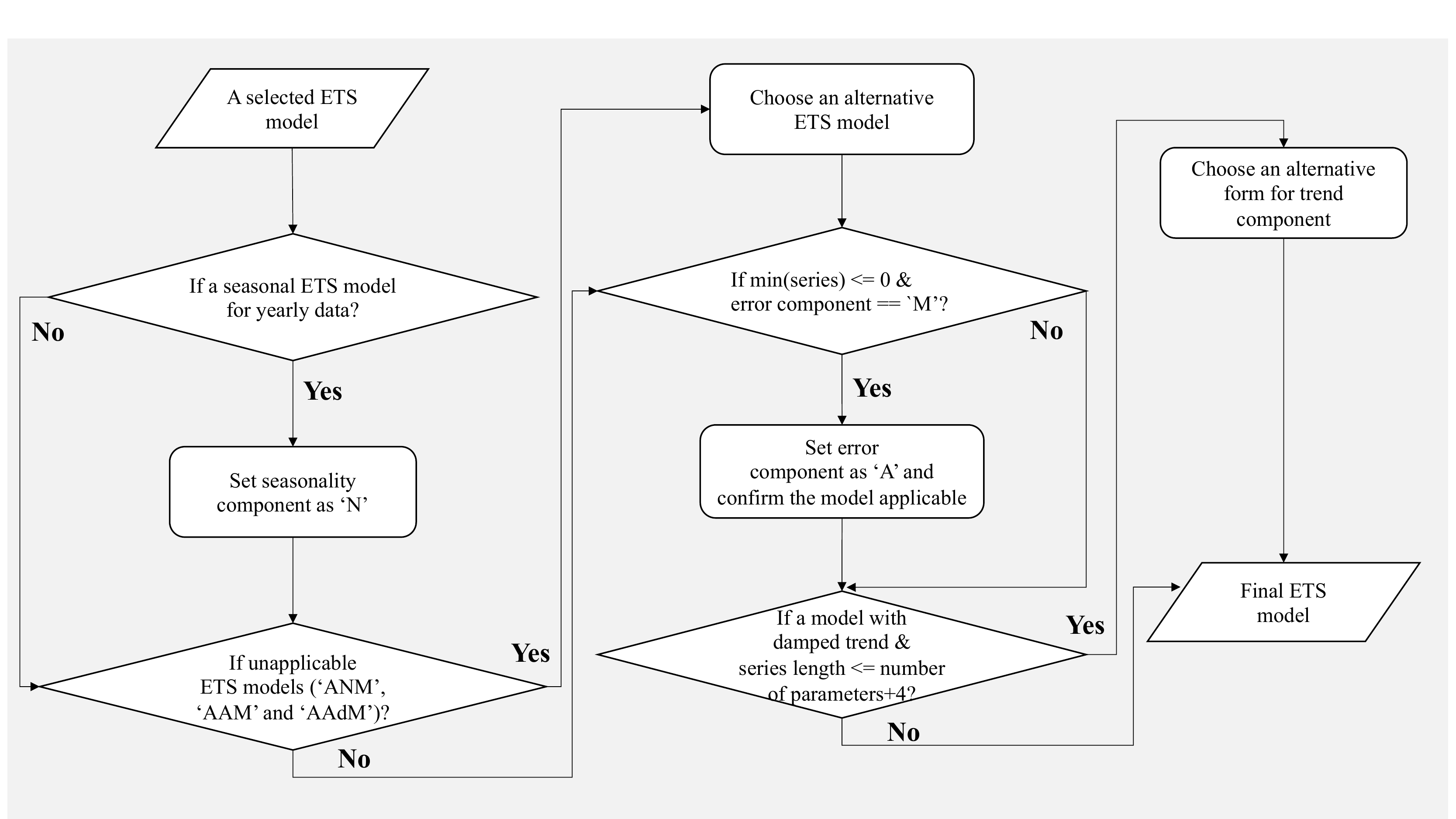}
  \caption{The procedure of model check and adjustment.}
  \label{model-check}
\end{figure}

Each pre-trained classifier will provide a certain probability that each component form is selected. Before fitting a selected model to data, it is required to confirm if the model from the combination of three predicted component forms is applicable. Fig.~\ref{model-check} presents the specific procedure of model check as well as the corresponding adjustment. 
\begin{itemize}
\tightlist
	\item \textbf{Check 1.} Due to that we train each classifier on all simulated data with different frequencies, it is unavoidable to mistakenly predict a seasonal ETS model for yearly data. If so, we set the seasonality component of the selected model as `N'.
	
	\item \textbf{Check 2.} If a selected model is unapplicable (like `ANM', `AAM', and `AAdM'), we replace its error and seasonality component forms with those from a model with the largest probability of being selected among all alternative models. The selected probability of an alternative model is calculated by multiplying the probabilities of each component.
	
	\item \textbf{Check 3.} Applying a selected model with multiplicative errors to a time series with zero or negative values would lead to instability of the model \citep{hyndman2018forecasting}. To tackle this issue, we change its error component form to `A'. Subsequently, we need to check if the updated model is applicable. If not, we replace its seasonality component form with one with the largest probability among all alternative forms.
	
	\item \textbf{Check 4.} To ensure a valid estimation of parameters in a model with a damped trend, the function \code{forecast::ets()} \citep{Rforecast} requires that the number of a time series should be larger than the number of parameters plus four when fitting it to data. If this condition is not met, we change a damped trend form to another one with a selected probability being ranked second among all trend component forms.

\end{itemize}

\FloatBarrier

\section{Empirical evaluation}
\label{experiments}

\subsection{Simulation study}
\subsubsection{Data simulation}
Considering that M4 competition data \citep{Makridakis2020-mm} has good representativeness of reality \citep{spiliotis2020forecasting}, to simulate some series that are realistic, we obtain some prior information from M4 data for this simulation study, as shown in Table~\ref{tab:SimP}.
The length sets for simulating time series with different frequencies are obtained from M4 data set. 
Particularly, for an ETS model with `Ad' trend, it is required to set $phi$ parameter to control the degree of damping, and we let $phi \sim U(0.80, 0.98)$ with reference to the function \code{forecast::ets()} \citep{Rforecast}. A model set with 15 applicable ETS models is used for quarterly and monthly data simulation, while another one with six models is for yearly data. In simulating a time series with a specific frequency, we randomly chose an ETS model form as well as a length value from the corresponding model set and length set. 

\begin{table}[htbp]
  \centering
  \caption{Parameter settings for time series simulation using \code{smooth::sim.es()} \citep{smooth}.}
  	\scalebox{0.95}{
  \begin{tabular}{p{0.1\columnwidth}p{0.35\columnwidth}p{0.5\columnwidth}}
    \toprule
    Parameter & \multicolumn{1}{l}{Description} & \multicolumn{1}{l}{Value}\\
    \midrule
    \code{frequency} & Period of time series & 1: yearly; 4: quarterly; 12: monthly\\
    \code{model} &The 15 applicable models, six for yearly data &Randomly chosen from the model sets\\
    \code{obs} & Length of time series &Randomly chosen from the length sets obtained from M4 data set  \\
    \code{phi} & Damping parameter for a model with `Ad' trend &U(0.80, 0.98) \\
    \bottomrule
    \end{tabular}
    }
  \label{tab:SimP}
\end{table}
\FloatBarrier

\begin{figure}[h!]
  \centering
  \includegraphics[width=12cm,height=16cm]{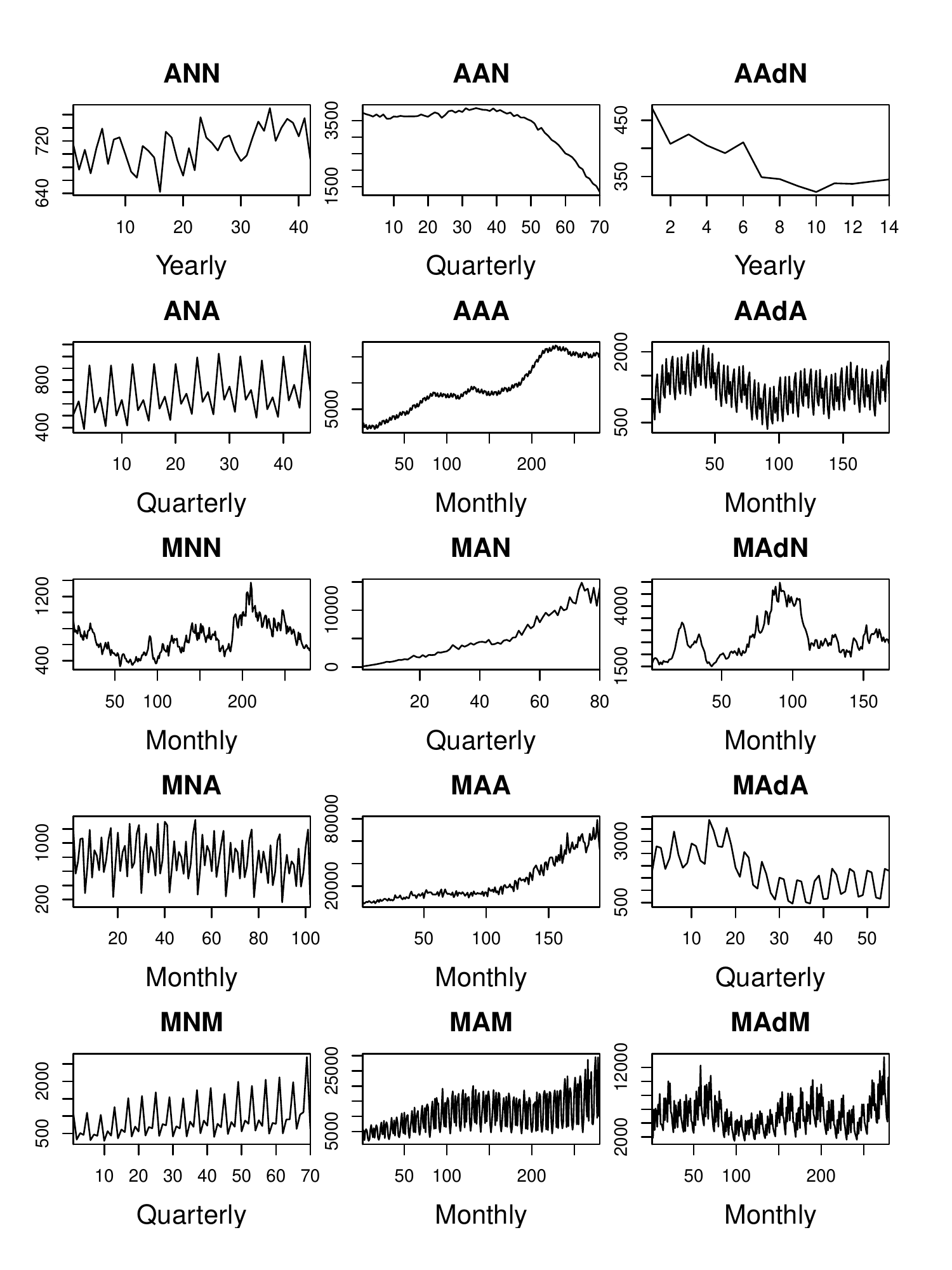}
  \caption{Typical simulated series from the 15 models in the ETS method.}
  \label{typical-series}
\end{figure}

Finally, we simulated 60,000 yearly, 60,000 quarterly and 120,000 monthly time series. The numbers for each seasonal and non-seasonal ETS model are approximately 12,000 and 22,000.
Fig.~\ref{typical-series} shows some typical simulated time series from the 15 applicable models.

\begin{figure}[h!]
  \centering
  \includegraphics[width=1\linewidth]{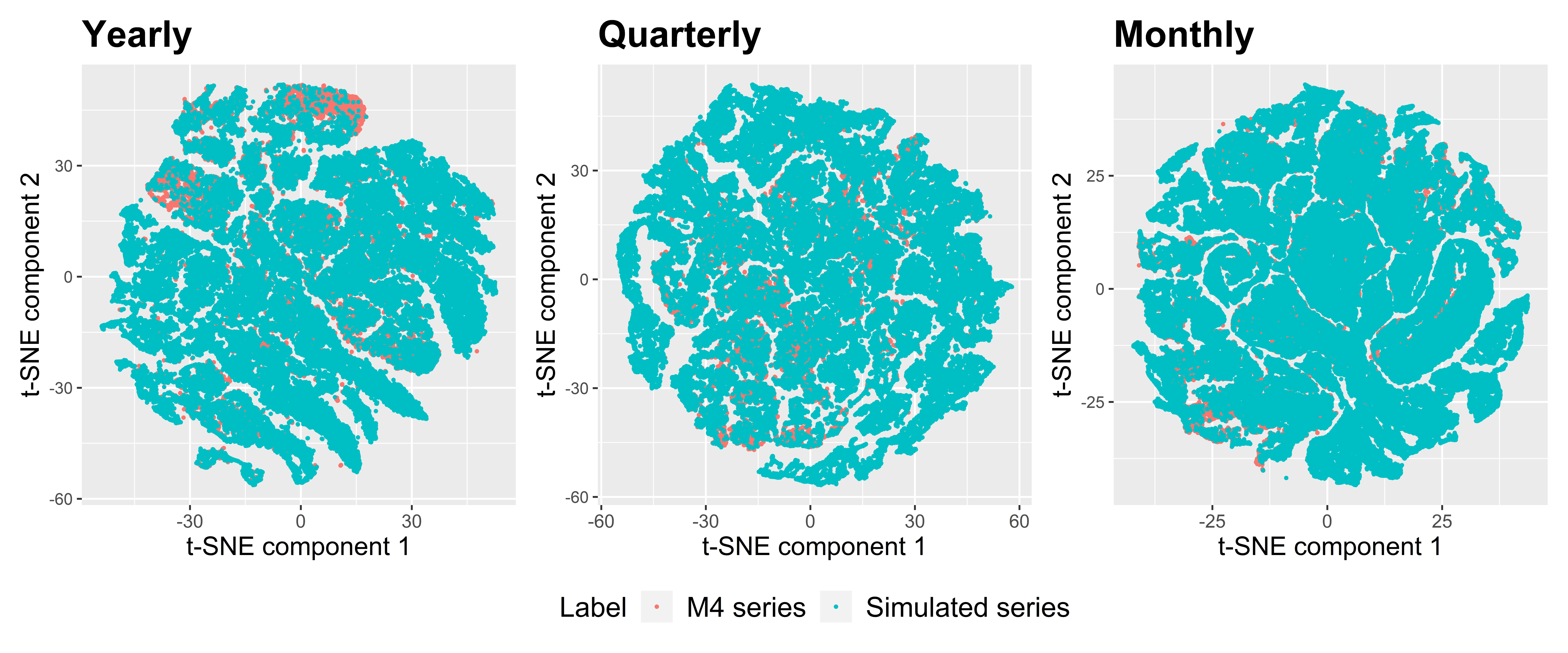}
  \caption{M4 and simulated data sets in a two-dimensional feature space using t-SNE.}
  \label{instance-space}
\end{figure}

To visualize if the simulated data set has a good representation of reality, we project the simulated data as well as M4 data into a two-dimensional feature space with t-SNE \citep{van2008visualizing}, as shown in Fig.~\ref{instance-space}.
Given the two-dimensional feature spaces of M4 and simulated data sets, we quantify the miscoverage of the simulated data set over M4 data set in the following steps \citep{kang2020gratis}.
 \begin{itemize}
 \tightlist
\item Cut the x and y dimensions of the instance space into $N_{b}=30$ bins, and obtain $N_{b}^{2}=900$ subgrids.
\item In the constructed subgrids, let
\begin{equation}
\mathcal{I}_{i, \mathrm{~S}}=\left\{\begin{array}{l}
0 \quad \text { if no time series in the simulated data set fall into the $i$-th subgrid;} \\
1 \quad \text { otherwise. }
\end{array}\right.
\end{equation}
An analogous definition of $\mathcal{I}_{i, \mathrm{~M}}$ for M4 data set.
\item The relative miscoverage of the simulated data set over M4 data set is
\begin{equation}
\text {miscoverage}_{\mathrm{S} / \mathrm{M}}=N_{b}^{-2} \sum_{i=1}^{N_{b}^{2}}\left[\left(1-\mathcal{I}_{i, \mathrm{~S}}\right) \times \mathcal{I}_{i, \mathrm{~M}}\right].
\end{equation}

\end{itemize}
 The miscoverage values of simulated data over M4 data are 0.026, 0.003 and 0.002 for yearly, quarterly and monthly data respectively. Together with Fig.~\ref{instance-space}, we conclude that the simulated data set has good coverage over M4 data, and thus has good representativeness of reality.

 \subsubsection{Performance of model selection on the simulated series}
We split the simulated data into two parts: 80\% of the series is used for training data and the rest for testing data. Three sets of optimal hyper-parameters of LightGBM from a search in subspaces of the hyper-parameter spaces are obtained by using a 5-fold cross-validation procedure on the training data. More details about these optimal hyper-parameters are available in Appendix~\ref{setup}.

In this simulation study, to demonstrate the model selection ability of the proposed method, we compare its performance on the testing data with that of the information criteria method. 
Two widely used metrics Accuracy and Macro $\mathrm{F_{1}}$-score are used for the evaluation of classification. The Accuracy is calculated as 
\begin{equation}
\mathrm{{ Accuracy }}=\frac{TS}{N}\times 100\%,
\end{equation}
where $TS$ is the number of series with true selected component forms or models, and $N=48,000$ is the number of testing series.

The $\Fscore_{i}$ 
\begin{equation}
\mathrm{\Fscore_{i}}=2 \frac{ { Recall }_{i}  { Precision }_{i}}{ { Recall }_{i}+ { Precision }_{i}} \times 100\%
\end{equation}
is used to measure the classification performance of the $i$-th class by combining its Recall and Precision into a single metric.

The Macro $\mathrm{F_{1}}$-score
\begin{equation}
\mathrm{Macro \Fscore}=(\frac{1}{m} \sum_{i=1}^{m} F_{1}\mbox{-}score_{i}) \times 100\%
\end{equation}
gives the same importance to each class by equally averaging $\Fscore_{i},i=1,...,m$ of all classes where $m$ is the number of classes.

Table~\ref{model-selection-accuracy} presents the classification performance of our method against the information criteria method in the tasks of separate ETS model component selection as well as whole model selection. The ETS method using information criteria for model selection is implemented using the function \code{forecast::ets()} \citep{Rforecast} with AICc as a default model selection criterion.

\FloatBarrier

\FloatBarrier

 From Table~\ref{model-selection-accuracy}, we can observe that our proposed method is superior to the information criteria method in all tasks on the simulated data.

\begin{center}
\begin{table}[H]
	\normalsize
	\centering
	\caption{Classification performance of our proposed method against the information criteria method with regard to Accuracy and Macro $\Fscore$. Entries in \textbf{bold} highlight that our approach outperforms the information criteria approach.}
	\label{model-selection-accuracy}
		\begin{adjustbox}{width=0.8\textwidth}
		\begin{tabular}{llrr}
			\toprule
 &&Accuracy &Macro $\Fscore$ \\
\cmidrule(lr){3-4}

\multirow{2}{*}{Error component} & Information criteria &85.05\% &87.29\% \\
 & Feature-based &\textbf{88.35\%} & \textbf{90.58\%} \\
 \cmidrule(lr){3-4}
\multirow{2}{*}{Trend component} & Information criteria & 69.73\% & 68.54\% \\
 & Feature-based &\textbf{74.41\%}&\textbf{74.32\%} \\
 \cmidrule(lr){3-4}
\multirow{2}{*}{Seasonality component} & Information criteria&92.27\% &{88.77\%} \\
 & Feature-based &\textbf{95.48\%} & \textbf{94.08\%} \\
 \cmidrule(lr){3-4}
\multirow{2}{*}{ETS model} & Information criteria &55.50\% &{54.32\%} \\
 & Feature-based&\textbf{63.51\%}&\textbf{65.71\%}\\
			\midrule
		\end{tabular}
	\end{adjustbox}
\end{table}
\end{center}

 \subsubsection{Feature importance analysis for each component selection}
\begin{figure}[h!]
  \centering
  \includegraphics[width=0.9\linewidth]{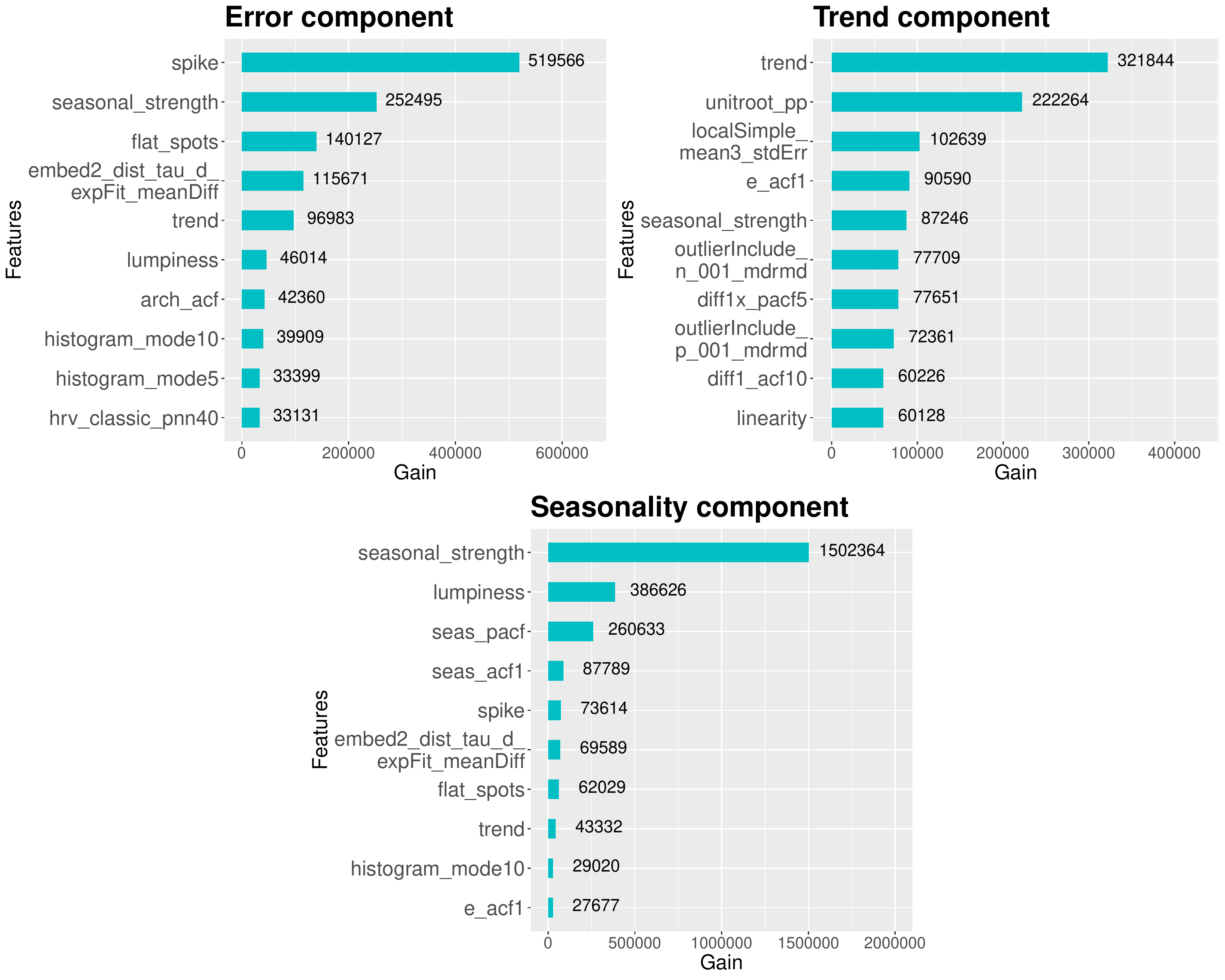}
  \caption{Feature importance for each component selection.}
  \label{feature-importance}
\end{figure}

To have a better understanding of which features are important to each component form prediction, we identify the ten most important features among 59 features by comparing their information gain \citep{ke2017lightgbm}. The more the gain is, the more important the feature is. Fig.~\ref{feature-importance} presents the feature importance results for each component form prediction. We can observe that

\begin{itemize}
\tightlist
	\item \textbf{Error component.} 
    The error related feature \texttt{spike} that calculates the variance of the leave-one-out variances of the remainder from an STL decomposition plays the most important role in the identification of the error component form. The feature \texttt{flat\_spots} that measures the longest flat spot in a series also has a significant effect on error component form identification. 
    What's more, the seasonal and trend strength of a series has an impact on error component selection in the sense that the error component is obtained after detrending and seasonal adjustment.
	\item \textbf{Trend component.} Some trend related features such as \texttt{trend} and \texttt{unitroot\_pp} are identified as the two most important features. The feature \texttt{localSimple\_mean3\_stdErr} reflecting mean error from a rolling forecasting is also important to the trend component selection.
		\item \textbf{Seasonality component.} Some seasonality related features such as \texttt{seasonal\_strength}, \texttt{seas\_pacf} and \texttt{seas\_acf1} are identified as important features. The feature \texttt{lumpiness} has a significant impact on the seasonality component selection due to that it calculates the variance of variances on different blocks with length of a seasonal period.
		
	\item \textbf{Summary.} The most important features are either from \pkg{M4metalearning} \citep{montero2020fforma} or \pkg{Rcatch22} \citep{Rcatch22}, which justifies the suitability of using two sets of features for the tasks of component selection.
\end{itemize}

\FloatBarrier

\subsection{Forecasting with M4 competition data}
To confirm the effectiveness of our proposed approach in real data, we compare our forecasts with those from the information criteria method over M4 data set. 
Considering that ETS models are not feasible for data with multiple seasonalities (e.g., daily and hourly data) or frequency larger than 24 (e.g., weekly data), we consider yearly, quarterly, and monthly subsets of M4 data set for this forecasting evaluation. The series in M4 competition come from different fields such as demographic, industry, finance, economics and others. The M4 dataset is in the \pkg{M4comp2018} \proglang{R} package~\citep{montero2018m4comp2018}. The forecasting horizons for yearly, quarterly and monthly data are 6, 8 and 18 respectively. The total number of series used in this forecasting evaluation is 23,000, 24,000 and 48,000 for yearly, quarterly and monthly data respectively. 

We employ two widely used metrics namely MASE (the Mean Absolute Scaled Error) \citep{hyndman2006another} and sMAPE (the Symmetric Mean Absolute Percentage Error) for point forecast evaluation. These two metrics are calculated as
\begin{equation*}
\begin{aligned}
\mathrm{MASE}&=\frac{1}{h}\frac{\sum_{t=1}^h\mid y_{T+t}-\hat{y}_{T+t} \mid}{\frac{1}{T-s}\sum_{t=s+1}^T\mid y_{t}-y_{t-s} \mid},\\
\mathrm{sMAPE}&=\frac{1}{h}\sum_{t=1}^h\frac{2\mid y_{T+t}-\hat{y}_{T+t}\mid}{\mid y_{T+t}\mid+\mid\hat{y}_{T+t} \mid},\\
\end{aligned}
\end{equation*}

where $\hat{y}_{t}$ is the
point forecast of the real value $y_{t}$ of the time series at point $t$, $h$ is the forecasting horizon, and $s$ is the seasonal period of data.

The MSIS (Mean Scaled Interval Score)~\citep{Gneiting2007a}  
\begin{equation*}
\resizebox{0.8\textwidth}{!} 
{$\text{MSIS}=\frac{1}{h}\frac{\sum_{t=1}^{h}\left(y_{T+t}^{u}-y_{T+t}^{l}\right)+\frac{2}{\alpha}\left(y_{T+t}^{l}-y_{T+t}\right) \mathbbm{1}_{\left\{y_{T+t}<y_{T+t}^{l}\right\}}+\frac{2}{\alpha}\left(y_{T+t}-y_{T+t}^{u}\right) \mathbbm{1}_{\left\{y_{T+t}>y_{T+t}^{u}\right\}}}{\frac{1}{T-s} \sum_{t=s+1}^{T}\left|y_{t}-y_{t-s}\right|}$}
\end{equation*}
is used to measure the accuracy of a $(1-\alpha) \times 100 \%$ prediction interval $\left(y_{T+t}^{l}, y_{T+t}^{u}\right)$ where $\mathbbm{1}_{A}$ is the indicator function for the condition $A$, and we set $\alpha=0.05$.

\subsection{Forecasting results}
The ETS method using information criteria for model selection is implemented using the function \code{forecast::ets()} \citep{Rforecast} with AICc as a default model selection criterion.
For our method, we use the pre-trained classifiers from the simulation study for ETS model selection, and then make predictions using these selected models. 
 Table~\ref{mean-accuracy} presents the mean forecasting results, and we can observe that:
\begin{itemize}
\tightlist
	\item \textbf{Point forecasts.} Our method resulted in better point forecasts at almost all the forecasting horizons for yearly (except for $h=1\mbox{-}4$ regarding sMAPE), quarterly (except for $h=7\mbox{-}8$ regarding MASE) and monthly data. 
	\item \textbf{Prediction intervals.} The performance of our approach is not better than that of the information criteria method regarding the mean of MSIS. 
\end{itemize}
\begin{center}
\begin{table}[H]
	\normalsize
	\centering
	\caption{Performance of our proposed method against the information criteria method with regard to the mean of MASE, sMAPE and MSIS values over M4 yearly, quarterly and monthly data. Entries in \textbf{bold} highlight that our approach outperforms the information criteria approach.}
	\label{mean-accuracy}
		\begin{adjustbox}{width=1\textwidth}
		\begin{tabular}{llrrrrrrrrrrrr}
			\toprule
&& \multicolumn{4}{c}{MASE}& \multicolumn{4}{c}{sMAPE}&\multicolumn{4}{c}{MSIS}\\
\cmidrule(lr){3-6} \cmidrule(lr){7-10} \cmidrule(lr){11-14}
& &1-2& 3-4 & 5-6& 1-6 & 1-2 & 3-4 & 5-6 & 1-6 & 1-2 & 3-4  & 5-6 & 1-6   \\
\cmidrule(lr){3-6} \cmidrule(lr){7-10} \cmidrule(lr){11-14}
\multirow{2}{*}{Yearly} & Information criteria & 1.934 & 3.456 & 4.943 & 3.444 & 9.809 & 15.743 & 20.516 & 15.356 &{18.983} &{34.772} &{50.934} &{34.897} \\
 & Feature-based & \textbf{1.931} & \textbf{3.392} & \textbf{4.820} & \textbf{3.381} & 9.933 & 15.746 & \textbf{20.368} & \textbf{15.349} & 21.956 & 36.878 & 53.386 & 37.407 \\
\cmidrule(lr){3-6} \cmidrule(lr){7-10} \cmidrule(lr){11-14}
& & 1-3& 4-6& 7-8 & 1-8 & 1-3 & 4-6 & 7-8 & 1-8 & 1-3 & 4-6 & 7-8  & 1-8 \\
\cmidrule(lr){3-6} \cmidrule(lr){7-10} \cmidrule(lr){11-14}
\multirow{2}{*}{Quarterly} & Information criteria & 0.774 & 1.256 & 1.598 & 1.161 & 7.454 & 11.023 & 13.448 & 10.291 &{6.113} &{10.188} &{13.356} &{9.452} \\
 & Feature-based & \textbf{0.772} & \textbf{1.256} &{1.599} & \textbf{1.160} & \textbf{7.402} & \textbf{10.943} & \textbf{13.410} & \textbf{10.232} & 6.390 & 10.694 & 14.045 & 9.918 \\
\cmidrule(lr){3-6} \cmidrule(lr){7-10} \cmidrule(lr){11-14}
 && 1-6 & 7-12 & 13-18 & 1-18 & 1-6 & 7-12 & 13-18 & 1-18 & 1-6 & 7-12 & 13-18 & 1-18\\
\cmidrule(lr){3-6} \cmidrule(lr){7-10} \cmidrule(lr){11-14}
\multirow{2}{*}{Monthly} & Information criteria & 0.650 & 0.967 & 1.227 & 0.948 & 9.933 & 13.637 & 17.005 & 13.525 &{5.152} &{8.552} &{11.188} &{8.297} \\
 & Feature-based & \textbf{0.639} & \textbf{0.959} & \textbf{1.207} & \textbf{0.935} & \textbf{9.687} & \textbf{13.385} & \textbf{16.466} & \textbf{13.179} & 5.352 & 8.773 & 11.491 & 8.539\\
			\midrule
		\end{tabular}
	\end{adjustbox}
\end{table}
\end{center}

We further provide the corresponding median values in Table~\ref{median-accuracy}. We can observe that our proposed method is superior to the information criteria method at almost all the forecasting horizons for yearly (except for $h=1\mbox{-}2$ regarding MSIS), quarterly  (except for  $h=1\mbox{-}3$ regarding sMAPE and $h=7\mbox{-}8$ regarding point forecasts) and monthly (except for $h=1\mbox{-}6$ regarding sMAPE) data. 

\begin{center}
\begin{table}[H]
	\normalsize
	\centering
	\caption{Performance of our proposed method against the information criteria method with regard to the median of MASE, sMAPE and MSIS values over M4 yearly, quarterly and monthly data.}
	\label{median-accuracy}
		\begin{adjustbox}{width=1\textwidth}
		\begin{tabular}{llrrrrrrrrrrrr}
			\toprule
&& \multicolumn{4}{c}{MASE}& \multicolumn{4}{c}{sMAPE}&\multicolumn{4}{c}{MSIS}\\
\cmidrule(lr){3-6} \cmidrule(lr){7-10} \cmidrule(lr){11-14}
& &1-2& 3-4 & 5-6& 1-6 & 1-2 & 3-4 & 5-6 & 1-6 & 1-2 & 3-4  & 5-6 & 1-6   \\
\cmidrule(lr){3-6} \cmidrule(lr){7-10} \cmidrule(lr){11-14}
\multirow{2}{*}{Yearly} & Information criteria & 1.339 & 2.255 & 3.162 & 2.329 & 5.342 & 8.669 & 11.629 & 8.966 & 7.777 & 13.905 & 20.623 & 15.487 \\
 & Feature-based & \textbf{1.318} & \textbf{2.214} & \textbf{3.102} & \textbf{2.289} & \textbf{5.263} & \textbf{8.426} & \textbf{11.320} & \textbf{8.721} &{8.238} & \textbf{13.162} & \textbf{18.278} & \textbf{15.280} \\

\cmidrule(lr){3-6} \cmidrule(lr){7-10} \cmidrule(lr){11-14}
& & 1-3& 4-6& 7-8 & 1-8 & 1-3 & 4-6 & 7-8 & 1-8 & 1-3 & 4-6 & 7-8  & 1-8 \\
\cmidrule(lr){3-6} \cmidrule(lr){7-10} \cmidrule(lr){11-14}
\multirow{2}{*}{Quarterly} & Information criteria & 0.573 & 0.910 &{1.123} & 0.886 &{3.513} & 5.598 & 6.835 & 5.608 & 3.993 & 6.199 & 7.658 & 5.977 \\
 & Feature-based & \textbf{0.570} & \textbf{0.904} & 1.124 & \textbf{0.886} &3.530& \textbf{5.597} &{6.882} & \textbf{5.600} & \textbf{3.980} & \textbf{6.115} & \textbf{7.575} & \textbf{5.939} \\

\cmidrule(lr){3-6} \cmidrule(lr){7-10} \cmidrule(lr){11-14}
 && 1-6 & 7-12 & 13-18 & 1-18 & 1-6 & 7-12 & 13-18 & 1-18 & 1-6 & 7-12 & 13-18 & 1-18\\
\cmidrule(lr){3-6} \cmidrule(lr){7-10} \cmidrule(lr){11-14}
\multirow{2}{*}{Monthly} & Information criteria & 0.497 & 0.708 & 0.877 & 0.736 & 4.223 & 6.654 & 8.329 & 6.995 & 3.225 & 4.933 & 6.339 & 5.040 \\
 & Feature-based & \textbf{0.491} & \textbf{0.705} & \textbf{0.867} & \textbf{0.728} &{4.247} & \textbf{6.641} & \textbf{8.258} & \textbf{6.964} & \textbf{3.219} & \textbf{4.917} & \textbf{6.260} & \textbf{5.030}\\

			\midrule
		\end{tabular}
	\end{adjustbox}
\end{table}
\end{center}

\subsection{Forecasting accuracy analysis}
\begin{figure}[h!]
  \centering
  \includegraphics[width=1\linewidth]{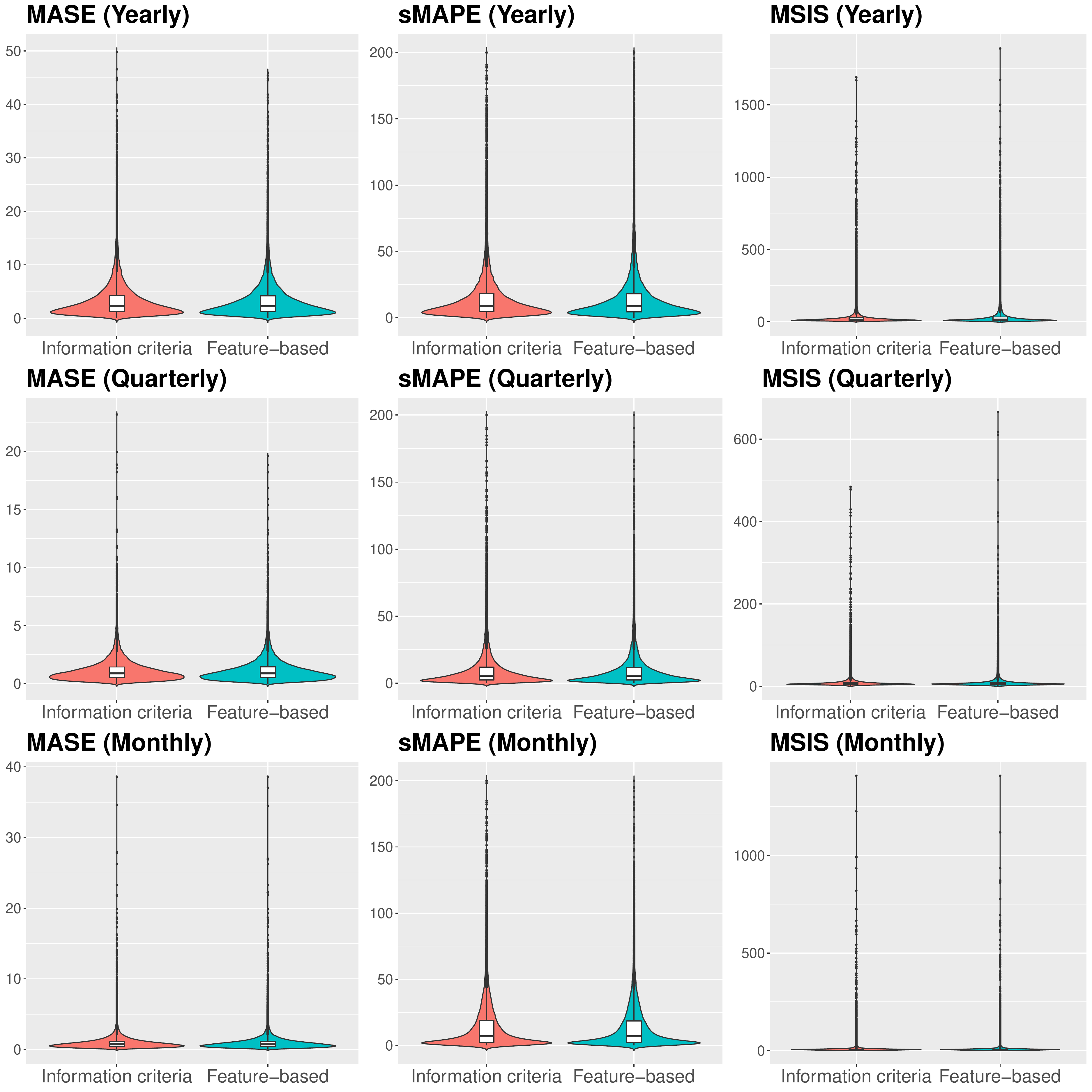}
  \caption{Violin plots of MASE, sMAPE and MSIS values over yearly, quarterly and monthly data}
  \label{plot-error}
\end{figure}

We use violin plots to visually show the distribution of MASE,  sMAPE and MSIS values at all horizons, as shown in Fig.~\ref{plot-error}.
We can observe that for yearly and quarterly data, the ranges of MASE of our method are smaller than that of the information criteria method. For monthly data, the distributions of forecasting accuracy of both methods are similar to each other. 

Together with Tables~\ref{mean-accuracy} and~\ref{median-accuracy} , we can conclude that the proposed method resulted in some unusual forecasts that have an effect on the mean of MSIS values, as shown in the first and second plots in the third column of Fig.~\ref{plot-error}. 
We find that these outliers come from some unusual series that let our method select some models with multiplicative errors for them, leading to large variances of prediction errors, and thus poor prediction intervals. Some of these unusual series can be available at \url{https://github.com/Richard759/fETSmcs/blob/master/bad_case.pdf}.

\subsection{Significance tests}
 We perform Diebold-Mariano (DM) tests \citep{harvey1997testing} to verify if forecasts from the proposed approach are significantly better or worse than those of the information criteria method.
The null hypothesis of this test is that the two approaches have the same forecast accuracy. The alternative hypothesis is that our method is less or more accurate than the benchmark method. In these tests, if the DM tests statistic falls in the lower or upper 2.5\% tail of the standard normal distribution, then we reject the null hypothesis and accept the alternative hypothesis. The DM tests are implemented using \code{forecast::dm.test()} \citep{Rforecast} in \proglang{R}. 
 
From Table~\ref{dm-tests}, we can observe that:
\begin{itemize}
\tightlist
	\item For yearly data, our approach significantly outperforms the information criteria method at long-term ($h=5\mbox{-}6$) and  all ($h=1\mbox{-}6$) horizons. 
	\item For quarterly data,  our approach is significantly superior to the information criteria method at mid ($h=4\mbox{-}6$), long-term  ($h=7\mbox{-}8$) and all ($h=1 \mbox{-}8$) horizons.
	\item For monthly data, our approach is significantly better than the information criteria method at long-term ($h=13\mbox{-}18$) and all ($h=1\mbox{-}18$) horizons. 
	\item The proposed method resulted in robust forecasts at long-term and all horizons for all data.
 
\end{itemize}

\begin{center}
\begin{table}[H]
	\normalsize
	\centering
	\caption{Diebold-Mariano (DM) tests for comparing the predictive accuracy of the proposed method with the information criteria method.
	The entries show the percentage of times the forecasts of our method are significantly better or worse than the forecasts of the information criteria method at different horizons. Entries in \textbf{bold} highlight that our method is significantly better than the information criteria method. }
	\label{dm-tests}
		\begin{adjustbox}{width=1\textwidth}
		\begin{tabular}{lrrrrrrrrrrrr}
			\toprule
& \multicolumn{4}{c}{Yearly} & \multicolumn{4}{c}{Quarterly} & \multicolumn{4}{c}{Monthly}                                                           \\
\cmidrule(lr){2-5} \cmidrule(lr){6-9} \cmidrule(lr){10-13}
& 1-2           & 3-4           & 5-6           & 1-6            & 1-3           & 4-6           & 7-8           & 1-8           & 1-6           & 7-12          & 13-18         & 1-18            \\
\cmidrule(lr){2-5} \cmidrule(lr){6-9} \cmidrule(lr){10-13}
better&1.600 & 3.574 & \textbf{5.770} & \textbf{16.839} & 2.350 & \textbf{9.163} & \textbf{6.329} & \textbf{17.879} & 11.135 & 16.683 & \textbf{20.410} & \textbf{22.846} \\
worse&1.630 & 3.613 & 5.448 & 15.539 & 2.646 & 8.921 & 6.096 & 17.242 & 11.279 & 17.031 & 19.779 & 22.238\\
			\midrule
		\end{tabular}
	\end{adjustbox}
\end{table}
\end{center}

\section{Case study: forecasting with a monthly hospital data set}
\label{case-study}
To further demonstrate the practical value of the proposed method in even specific forecasting domains, we apply our method to a monthly hospital data set. The data set consisting of 767 monthly series has been analyzed by \cite{hyndman2008forecasting}. The data set is in \pkg{expsmooth} \proglang{R} package~\citep{expsmooth}.
We save the last 18 observations for forecasting evaluation.
We employ the pre-trained classifiers from the simulation study to select an appropriate ETS model for any series in this set, and then make predictions using these selected models. From Table~\ref{mean-accuracy-hospital}, we can find that our proposed approach resulted in better forecasting results than the information criteria method at almost all horizons (except for $h=1\mbox{-}6$ regarding MSIS) in terms of point forecasts and prediction intervals.

\begin{center}
\begin{table}[H]
	\normalsize
	\centering
	\caption{Performance of our proposed method against the information criteria method with regard to the mean of MASE, sMAPE and MSIS values over the hospital data. Entries in \textbf{bold} highlight that our approach outperforms the information criteria approach.}
	\label{mean-accuracy-hospital}
		\begin{adjustbox}{width=1\textwidth}
		\begin{tabular}{lrrrrrrrrrrrr}
			\toprule
& \multicolumn{4}{c}{MASE}& \multicolumn{4}{c}{sMAPE}&\multicolumn{4}{c}{MSIS}\\
\cmidrule(lr){2-5} \cmidrule(lr){6-9} \cmidrule(lr){10-13}
& 1-6 & 7-12 & 13-18 & 1-18 & 1-6 & 7-12 & 13-18 & 1-18 & 1-6 & 7-12 & 13-18 & 1-18\\
\cmidrule(lr){2-5} \cmidrule(lr){6-9} \cmidrule(lr){10-13}
 Information criteria & 0.726 & 0.804 & 0.895 & 0.808 & 16.724 & 18.138 & 19.855 & 18.239 & 4.536 & 5.519 & 6.486 & 5.514 \\
  Feature-based & \textbf{0.712} & \textbf{0.780} & \textbf{0.865} & \textbf{0.786} & \textbf{16.364} & \textbf{17.778} & \textbf{19.353} & \textbf{17.832} & 4.556 & \textbf{5.433} & \textbf{6.422} & \textbf{5.470}\\
			\midrule
		\end{tabular}
	\end{adjustbox}
\end{table}
\end{center}

\section{Computational time for producing forecasts}
 We save and compare the computational time for producing forecasts from our method as well as the information criteria method over the above M4 and hospital data sets.  To ensure a fair comparison, all the experiments were carried out on an AMD Ryzen 7 4800H 2.90 GHz server with 8 cores. The computation time of the proposed method in the training phase for time series simulation, feature extraction and classifier building is 1,162, 1,683 and  7,386 seconds respectively. 
 
 From Table~\ref{time-cost}, we can observe that our proposed method is significantly faster than the information criteria method for producing forecasts in the online phase. It cannot be denied that the most computational process of our method is time series simulation, feature extraction, and classifier training in the training phase. However, this is acceptable in real application scenarios because classifiers only need to be trained once, and can be applied to different real data without training each time.

\begin{center}
\begin{table}[H]
	\normalsize
	\centering
	\caption{Computational time in seconds for producing forecasts from our method and information criteria method in the online phase.}
	\label{time-cost}
		\begin{adjustbox}{width=1\textwidth}
		\begin{tabular}{lrrrr}
			\toprule
 & \multicolumn{2}{c}{M4} & \multicolumn{2}{c}{Hospital} \\
\cmidrule(lr){2-3} \cmidrule(lr){4-5}
 & Information criteria & Feature-based & Information criteria & Feature-based \\
\cmidrule(lr){2-3} \cmidrule(lr){4-5}
Feature extraction &-  & 588 & - & 5 \\
Model selection and forecasting &5,782  & 367 & 67 & 4.1 \\

\midrule
Total & 5,782 & 955 & 67 & 9.1\\
			\midrule
		\end{tabular}
	\end{adjustbox}
\end{table}
\end{center}

\section{Discussions}
\label{discussion}
Considering that the existing ETS method using information criteria for model selection suffers from computational complexity when applied to large-scale time series data, we propose an efficient approach for ETS model selection that trains three classifiers on simulated data to predict model component forms. 
A simulation study has shown the model selection ability of the proposed method. DM tests results confirm that our approach produces robust forecasts at long-term and all horizons for yearly, monthly and quarterly data.  

The most computational process of the proposed method is time series simulation, feature extraction, and classifier building in the training phase. However, this is acceptable in practice because classifiers for component form prediction only need to be trained once, and can be applied to different real data without retraining.  The good forecasting results on a hospital data set from a specific domain have demonstrated the robustness of the proposed method.
We also have shown the proposed method is more efficient to produce forecasts over the M4 and hospital data sets in the online phase.

However, the good forecasting performance of the proposed approach depends on the selection of an appropriate set of time series features and careful parameter tuning. Compared with the information criteria method, our method achieved impressive performance, but sacrifices interpretability, that is, we have no idea about how well a selected model from our classifiers fits the data and how complicated the model is.

Our approach feature-based ETS model component selection (fETSmcs) is implemented as an \proglang{R} package. Our code is open-source and publicly available at \url{https://github.com/Richard759/fETSmcs}.

\section{Concluding remarks}
\label{conclusion}
To tackle the inefficiency issue in the current big data settings, we explore the use of features for ETS model selection by training classifiers on simulated data to predict component forms. We have provided a simulation study as well as two real data applications to demonstrate of the effectiveness the proposed method.

However, the proposed approach depends on the manual choice of an appropriate set of features. Some research \citep[e.g.,][]{li2020forecasting} introduces automatic features for forecast model combination, and one potential direction for this research would be to adopt automatic features for ETS model selection. A potential avenue for future research would be to design a model selection criterion that incorporates information criteria as well as feature-based selection, aiming at achieving impressive performance while still maintaining good interpretability.

\section{Acknowledgements}
We are grateful to the Editor and two anonymous reviewers for their helpful comments that improved the contents of this paper. 

We are also grateful to Professor Yanfei Kang from Beihang University for her feedback in the early stages of the development of the proposed method in this paper.

\bibliographystyle{model5-names}
\bibliography{fETSmcs}

\newpage
\begin{appendices}
\section{Feature descriptions}
Tables~\ref{tab:features1} and \ref{tab:features2} introduces the 59 used features as well as their corresponding descriptions.
\label{features}

\section{Experimental setup for LightGBM}
\label{setup}
We performed hyper-parameter optimization with grid search by measuring the AUC (Area Under the ROC (Receiver Operating Characteristic) Curve) on a 20\% holdout version of the training data in a 5-fold cross-validation procedure. The searching ranges of four hyper-parameters are as follows.
\begin{itemize}
  \tightlist

\item \texttt{num\_leaves}: The number of leaves ranges from $8$ to 128.
\item \texttt{min\_delta\_in\_leaf}: The minimum number of the records a leaf may have ranges from $10$ to $110$.
\item \texttt{max\_bin}: The maximum number of bins that feature values will be bucketed in ranges from $5$ to $255$.
\item \texttt{num\_boost\_round}: The number of boosting stages ranges from $100$ to $1100$.
\end{itemize}
Considering computation time, we set fixed values for some other hyper-parameters.
Table~\ref{param_clasifiers} reports the optimal values of four hyper-parameters of LightGBM as well as other hyper-parameters with fixed values for the three classifiers  $f_{e}$, $f_{t}$  and $f_{s}$.

 \begin{table}[htbp]
  \centering
  \caption{The \pkg{M4metalearning} features \citep{montero2020fforma} used for ETS model component selection.}
  	\label{alg:classification}
  		\scalebox{0.85}{
\begin{tabular}{p{0.05\columnwidth}p{0.18\columnwidth}p{0.77\columnwidth}}
    \toprule
    Index &Feature&Description\\
    \midrule
    $1$&x\_acf &First ACF of a series\\
     $2$&x\_acf10 &Sum of the squared first ten ACF of a series\\
     $3$&diff1\_acf1&First ACF of a differenced series\\
    $4$&diff1\_acf10 &Sum of the squared first ten ACF of a differenced series\\
     $5$&diff2\_acf1 &First ACF of a twice-differenced series\\
     $6$&diff2\_acf10&Sum of squared fist ten ACF of a series\\
         $7$&seas\_acf1 &ACF of a seasonally differenced series, 0 for non seasonal series\\
     $8$&ARCH.LM  &A statistic for AR conditional heteroscedasticity\\
     $9$&crossing\_point&The number of times a series crosses its median\\
         $10$&entropy &The spectral entropy of a series\\
     $11$&flat\_spots &The number of flat spots in a series\\
     $12$&arch\_acf&Sum of squares of first 12 ACF of a pre-whitened series\\
     $13$&garch\_acf &Sum of squares of first 12 ACF of the squared residuals from a fitted GARCH(1,1) model\\
     $14$&arch\_r2&$R^{2}$ value of an AR model applied to a pre-whitened series\\
     $15$&garch\_r2 &$R^{2}$ value of a GARCH(1,1) model applied to a pre-whitened series \\
     $16$&hurst&Hurst\\
     $17$&lumpiness&Lumpiness\\
     $18$&nonlinearity&Nonlinearity\\
    $19$&x\_pacf5 &Sum of squared first 5 PACF of a series\\
     $20$&diff1x\_pacf5 &Sum of squared first 5 PACF of a differenced series\\
     $21$&diff2x\_pacf5&Sum of squared first 5 PACF of a twice differenced series\\
         $22$&seas\_pacf&PACF of at first seasonal lag, 0 for non seasonal series\\
     $23$&nperiods&The number of seasonal periods in a series\\
     $24$&seasonal\_period & The length of the seasonal period\\
     $25$&trend&Trend\\
    $26$&spike&Spike\\
     $27$&linearity&Linearity\\
     $28$&curvature&Curvature\\
    $29$&e\_acf1&First ACF of the remainder series from an STL decomposition\\
     $30$&e\_acf10 &Sum of first 10 squared ACF of the remainder series from an STL decomposition\\
     $31$&seasonal\_strength&Strength of seasonality\\
     $32$&peak &Peak\\
     $33$&trough &Trough\\
     $34$&stability&Stability\\
         $35$&unitRoot\_kpss & A statistic for unit root test\\
     $36$&unitRoot\_pp &Another statistic for unit root test\\
     $37$&series\_length&Series length\\
    \bottomrule
    \end{tabular}
    }
  \label{tab:features1}
\end{table}
\FloatBarrier

 \begin{table}[htbp]
  \centering
  \caption{The \pkg{Rcatch22} features \citep{lubba2019catch22} used for ETS model component selection.}
  	\label{alg:classification}
  		\scalebox{0.9}{
\begin{tabular}{p{0.05\columnwidth}p{0.42\columnwidth}p{0.53\columnwidth}}
    \toprule
    Index &Feature&Description\\
    \midrule
     $38$&histogram\_mode5 &Mode of z-scored distribution using 5-bin histogram\\
     $39$&histogram\_mode10&Mode of z-scored distribution using 10-bin histogram\\
     $40$&binaryStats\_mean\_longStretch1&Longest period of consecutive values above mean\\
     $41$&outlierInclude\_p\_001\_mdrmd&Time intervals between extreme events above mean\\
    $42$&outlierInclude\_n\_001\_mdrmd &Time intervals between extreme events below mean\\
     $43$&f1ecac&First 1/e crossing of ACF\\
     $44$&firstmin\_ac&First minimum of ACF\\
         $45$&summaries\_welch\_rect\_area\_5\_1&Total power of lowest fifth of frequencies in a power spectrum\\
              $46$&summaries\_welch\_rect\_centroid&Centroid of the Fourier power spectrum\\
     $47$&localSimple\_mean3\_stdErr&Prediction error from a rolling forecasting\\
     $48$&trev\_1\_num&A statistic for time-reversibility\\
     $49$&histogramAMI\_even\_2\_5&Automutual information\\
    $50$&autoMutualInfoStats\_40\_gaussian\_fmmi &First minimum of an information function\\
     $51$&hrv\_classic\_pnn40&Proportion of continuous differences above 0.04$\sigma$ \ \citep{mietus2002pnnx}\\
     $52$&binaryStats\_diff\_longStretch0&Longest length of incremental decreases\\
         $53$&motifThree\_quantile\_hh&Entropy of two continuous letters in a symbolization\\
         
     $54$&localSimple\_mean1\_tauresrat&Correlation changes after iterative differencing\\
     $55$&embed2\_dist\_tau\_d\_expFit\_meanDiff&An exponential fit to continuous  distances in 2-d space\\
     $56$&fluctAnal\_2\_dfa\_50\_1\_2\_logi\_prop\_r1&Proportion of slower timescale fluctuations with DFA scaling\\
    $57$&fluctAnal\_2\_rsrangefit\_50\_1\_logi\_prop\_r1 &Proportion of slower timescale fluctuations with rescaled range fits \\
     $58$&transitionMatrix\_3ac\_sumDiagCov&Trace of covariance of a transition matrix\\
     $59$&periodicityWang\_th0\_01&A periodicity measure \citep{wang2007structure}\\
    \bottomrule
    \end{tabular}
    }
  \label{tab:features2}
\end{table}
\FloatBarrier

\begin{table}[H]
  \centering
  \caption{Values of hyper-parameters of LightGBM on the training data.}
  \label{param_clasifiers}
	\scalebox{1}{
    \begin{tabular}{lrrr}
      \toprule
      Hyper-parameters & $f_{e}$   &$f_{t}$ & $f_{s}$ \\
      \midrule 
      \texttt{eta} & 0.05 & 0.05 & 0.05 \\
      \texttt{num\_leaves} & 92 & 80 & 64 \\
      \texttt{min\_data\_in\_leaf} & 90 & 100 & 60\\
      \texttt{max\_bin} & 175 & 225 & 175 \\
      \texttt{num\_boost\_round} & 600 &800 & 1000 \\
      \texttt{bagging\_fraction} & 0.8 & 0.8 & 0.8 \\
      \texttt{bagging\_freq} & 4 & 4 & 4 \\
      \texttt{feature\_fraction} & 0.7 & 0.7 & 0.7 \\
      \texttt{force\_col\_wise} & True & True & True \\
      \texttt{bagging\_seed} & 123 & 123 & 123 \\
      \bottomrule
    \end{tabular}
    }
 
\end{table}
\FloatBarrier

\end{appendices}

\end{document}